\documentclass{amsart}
\usepackage{amsmath,amsfonts,amssymb,amsthm}
\usepackage[english]{babel}
\usepackage{graphicx}
\usepackage{url}
\usepackage{palatino}
\usepackage{wrapfig}
\usepackage[hidelinks]{hyperref}
\usepackage[round]{natbib}
\usepackage[labelfont=bf,labelsep=period,font=small]{caption} 
\newcommand{\forarxiv}[1]{#1}
\newcommand{\notforarxiv}[1]{}

\newcommand{\eat}[1]{}

\begin{document}

\notforarxiv{
\begin{flushright}
Version dated: \today
\end{flushright}
\bigskip
\noindent RH: PHYLOGENETICS AND THE HUMAN MICROBIOME
\bigskip
\medskip
\begin{center}

\noindent{\Large \bf Phylogenetics and the human microbiome.}
\bigskip

\noindent {\normalsize \sc
Frederick A. Matsen IV$^1$}\\
\noindent {\small \it
$^1$
Program in Computational Biology, Fred Hutchinson Cancer Research Center, Seattle, WA, 91802, USA}\\
\end{center}
\medskip
\noindent{\bf Corresponding author:} Frederick A Matsen, Program in Computational Biology, Fred Hutchinson Cancer Research Center, Seattle, WA, 91802, USA; E-mail: matsen@fhcrc.org.\\
\vspace{1in}
}

\forarxiv{\
\title{Phylogenetics and the human microbiome.}
\author{Frederick A. Matsen IV}
\date{\today}
\begin{abstract}
}
\notforarxiv{
\newpage
\section{Abstract}
}

The human microbiome is the ensemble of genes in the microbes that live inside and on the surface of humans.
Because microbial sequencing information is now much easier to come by than phenotypic information, there has been an explosion of sequencing and genetic analysis of microbiome samples.
Much of the analytical work for these sequences involves phylogenetics, at least indirectly, but methodology has developed in a somewhat different direction than for other applications of phylogenetics.
In this paper I review the field and its methods from the perspective of a phylogeneticist, as well as describing current challenges for phylogenetics coming from this type of work.

\forarxiv{
\end{abstract}
\maketitle
}

\notforarxiv{
\vspace{4.5in}
\noindent (Keywords: human microbiome; human microbiota; microbial ecology; phylogenetic methods; 16S; metagenome)\\
\newpage
}

\section{Introduction}

The parameter regime and focus of human-associated microbial research sits outside of the traditional setting for phylogenetics methods development and application; why should our community be interested in what microbial ecologists and medical researchers have done?
The answer is simple: this system is data- and question-rich.
Microbes are now primarily identified by their molecular sequences because such molecular identification is much more straightforward to do in high throughput than morphological or phenotypic characterization.
Indeed, microbial ecology has recently become for the most part the study of the relative abundances of various sequences derived from the environment, even if the framework for understanding between-microbe relationships includes metabolic information and other information not derived directly from sampled molecular sequences.

Although there is something of a divide between phylogeny as practiced as part of microbial ecology on one hand and that for multicellular organisms on the other, there are many parallels between the two enterprises.
Both communities struggle with issues of sequence alignment, large-scale tree reconstruction, and species delimitation.
However, approaches differ between the microbial ecology community and that of eukaryotic phylogenetics, in part because the scope of the former contains an almost unlimited diversity of organisms, leading to additional problems above the usual.
The species concept is even more problematic for microbes than for multicellular organisms, and hence there is also considerable discussion concerning how to group them into species-like units.
Organizing microbes into a sensible taxonomy is a serious challenge, especially in the absence of obvious morphological features.

Because of this high level of diversity and challenges with species definitions, microbial ecology researchers have developed their own explicitly phylogenetic techniques for comparing samples rather than comparing on the level of species abundances.
Although there is some overlap with previous literature, these techniques could be used in a wider setting and may deserve broader consideration by the phylogenetics community.

The human-associated microbial assemblage is specifically interesting because questions of microbial genomics, translated into questions of function, have important consequences for human health.
Additionally, due to more than a century of hospital laboratory work, our knowledge about human-associated microbes is relatively rich.
This collection of microbes living inside and on our surfaces is called the \textit{human microbiota}, and the ensemble of genetic information in those microbes is called the \textit{human microbiome}, although usage of these terms varies \citep{Boon2013-mc}.

In this review I will describe phylogenetics-related research happening in microbial ecology and contrast approaches between microbial researchers and what I think of as the typical \emph{Systematic Biology} audience.
Despite an obvious oversimplification, I will use \textit{eukaryotic phylogenetics} to indicate what I think of as the mainstream of SB readership, and \textit{microbial phylogenetics} to denote the other.
I realize that there is substantial overlap--- for instance the microbial community is very interested in unicellular fungi, and additionally many in the SB community do work on microbes--- but this terminology will be useful for concreteness.
There is of course also substantial overlap in methodology, however as we will see there are significant differences in approach and the two areas have developed somewhat in parallel.
I will first briefly review the recent literature on the human microbiota, then describe novel ways in which human microbiome researchers have used trees.
I will finish with opportunities for the \textit{Systematic Biology} audience to contribute to this field.
I have made an explicit effort to make a neutral comparison between two directions rather than criticize the approximate methods common in microbial phylogenetics; indeed, microbial phylogenetics requires algorithms and ideas that work in parameter regimes an order of magnitude larger than typical for eukaryotic phylogenetics.

\section{The human microbiota}
The human microbiota is the collection of microbial organisms that live inside of and on the surface of humans.
These organisms are populous: it has been estimated that there are ten times as many bacteria associated with each individual than there are human cells of that individual.
The microbiota have remarkable metabolic potential, being an ensemble of genes estimated to be about 150 times larger than the human collection of genes \citep{qin2010human}.
Much of our metabolic interaction with the outside world is mediated by our microbiota, as it has important roles in immune system development, nutrition, and drug metabolism \citep{kau2011human,maurice2013xenobiotics}.
Our food and drug intake, in turn, impacts the diversity of microbes present.
Traditionally, our microbiota have been transmitted from mother to infant in the birth canal and by breastfeeding \citep[reviewed in][]{funkhouser2013mom}.
In this section I will briefly review what is known about the human microbiota and its effect on our health.

The human microbiota form an ecosystem.
It is dynamic in terms of taxonomic representation but apparently constant in terms of function \citep{hmp2012structure}.
There is a ``core'' microbiota which is shared between all humans \citep{turnbaugh2008core}.
The human microbiota is spatially organized, as can be seen on skin \citep{grice2009topographical}, with substantial variation in human body habitats across space and time \citep{costello2009bacterial}.
There is a substantial range of inter-individual versus intra-individual variation \citep{hmp2012structure}.

Our actions can shift the composition of our microbiota.
Changes in diet can very quickly shift its composition, but there is also a strong correlation between long-term diet and microbiota \citep{li2009human,wu2011linking}.
Antibiotics fundamentally disturb microbial communities, resulting in an effect that lasts for years \citep{jernberg2007long,dethlefsen2008pervasive,jakobsson2010short,dethlefsen2011incomplete}.

The microbiota interact on many levels with host phenotype \citep[reviewed in][]{cho2012human}.
The gut microbiota in particular correlates with health of individuals from the elderly in industrialized nations \citep{claesson2012gut} to children with acute metabolic dysfunction in rural Africa \citep{smith2013gut}.
Considerable attention has also been given to the interaction between gut microbiota and obesity, although the story is not yet clear.
An intervention study has established human gut microbes associated with obesity \citep{ley2006microbial}.
A causal role for the microbiota leading to obesity has been established for mice: an obese phenotype can be transferred from mouse to mouse by gut microbial transplantation \citep{turnbaugh2006obesity}, the pregnant human gut microbiota leads to obesity in mice \citep{koren2012host}, and probiotics can lead to a lean phenotype and healthy eating behavior \citep{poutahidis2013microbial}.
However, these promising leads have not yet been confirmed causally or in population studies of humans \citep{zhao2013gut}.
For example, a study of obesity in the old-order Amish did not find any correlation between obesity and particular gut communities \citep{zupancic2012analysis}.

Bacteria have been the primary focus of human microbiota research, and other domains have been investigated to a lesser extent.
Changes in archaeal and fungal populations have been shown to covary with bacterial residents \citep{hoffmann2013archaea} and have a nonuniform distribution across the human skin \citep{Findley2013-rm}.
Viral populations have been observed to be highly dynamic and variable across individuals \citep{reyes2010viruses,minot2011human,minot2013rapid}.
We will focus on bacteria here.

In this paper we will primarily be describing the human microbiota from a community-level phylogenetic perspective rather than from the fine-scale perspective of immune-mediated interactions between host and microbe \citep[reviewed in][]{hooper2012interactions}.
Our understanding of the true effect of the microbiota will eventually come from such a molecular-level understanding, although until we can characterize all of the molecular interactions between microbes and the human body, a broad perspective will continue to be important.

\section{Investigating the human microbiome via sequencing}
It is now possible to assay microbial communities in high throughput using sequencing.
One way is to amplify a specific gene in the genome for sequencing using polymerase chain reaction (PCR).
Scientists typically pick a ``marker'' gene in that case that is meant to recapitulate the ``overall'' evolutionary history of the microbes.
Another way is to randomly shear input DNA and/or RNA and then perform sequencing directly.
We will consistently refer to the former as a \textit{survey} and the second a \textit{metagenome}, although these words have not always been consistently used in the literature.

The Human Microbiome Project \citep{methe2012framework} generated lots of survey, metagenome, and whole-genome sequencing data and these data are available on a dedicated website\footnote{\url{http://www.hmpdacc.org/}}.
The MetaHIT study \citep{qin2010human} also generated lots of data but it is not available to outside researchers.

\subsection{Inferring microbial community composition using marker gene surveys}
Our modern knowledge of the microbial world is in a large part derived from the methods of Carl Woese and colleagues who pioneered the use of marker genes as a way to distinguish between microbial lineages \citep{fox1977comparative}.
Their work, and the scientists who followed them, focused on the 16S ribosomal gene (henceforth simply ``16S'') as a genetic marker.
This gene was chosen because it has regions of high and low diversity, which enable resolution on a variety of evolutionary time scales.
Regions of low diversity in 16S also enabled the development of the first ``universal'' 16S PCR primers \citep{lane1985rapid} which enabled detection of almost all bacteria and archaea regardless of whether they can be cultured.

In microbial ecology, the census of bacteria in a given environment using marker gene amplification and sequencing are generally called ``marker gene surveys.''
This terminology is equivalent to the ``barcoding'' terminology more commonly used for eukaryotic surveys using 18S or the fungal internal transcribed spacer (ITS).
Such surveys would ideally return a census of all of the microbes in a sample along with their abundances.

Where Woese and colleagues labored over digestion and gel electrophoresis to infer sequences, modern researchers have the luxury of high throughput sequencing.
This can be done with a high level of multiplexing, making an explicit trade-off between depth of sequencing for each specimen and the number of specimens able to be put on the sequencer at the same time.
This has led to extensive parallelization, most recently by sequencing dozens of samples at a time on the Illumina instrument \citep{degnan2011illumina,caporaso2012ultra}.
This brings up the question of how many sequences are needed to characterize the microbial diversity of a given environment.
To distinguish between two rather different samples, relatively few sequences per sample are required \citep{kuczynski2010microbial}, however, to compare more similar samples deeper sequencing is required.
In addition to sequencing samples across individuals, this parallelization has also enabled sampling through time \citep[e.g.][]{caporaso2011moving}.

Despite the high throughput and low cost of modern sequencing, inherent challenges remain for applications of marker gene sequencing to take a census of microbes.
Most fundamentally, various microbes have different DNA extraction efficiencies, even with stringent protocols, meaning that a collection of sequences need not be representative of the communities from which they were derived \citep{morgan2010metagenomic}.
Current high throughput sequencing technology is limited to a length that is shorter than most genes, which limits the resolution of the analyses.
``Primer bias,'' or differing amplification levels of various sequences based on their affinity for the primers \citep{suzuki1996bias,polz1998bias}, is a challenge and has led to the standardization of primers \citep{methe2012framework}.
Worse, multiplex PCR is known to create chimeric (i.e.\ spurious recombinant) sequences via partial PCR products \citep{hugenholtz2003chimeric,ashelford2005least,haas2011chimeric,schloss2011reducing}.
Correspondingly, chimera checking software has been developed \citep[including][]{ashelford2006new,edgar2011uchime}.
Also, 16S can be present in up to 15 copies and there can be diversity within the copies \citep{klappenbach2001rrndb}.
This can distort inferences concerning actual microbe abundances based on read copy number.
Recent work by \citet{kembel2012incorporating} implements the independent contrasts method \citep{felsenstein1985phylogenies} to correct for copy number, which has been helpful despite a moderate evolutionary signal in copy number variation \citep{klappenbach2000rrna}.
Some groups have reported advantages to using alternate single-copy genes as markers for characterization of microbial communities \citep[e.g.][]{mcnabb2004hsp65,case2007rpob}, however 16S remains the dominant locus used by a large margin.
A final cause of noise is next-generation sequencing error: this is certainly a problem for both surveys and metagenomes, but is becoming less of a problem as technology improves.
I will not address it specifically except in the inference of operational taxonomic units as described below.

\subsection{Metagenomes}
As described above ``metagenome" means that DNA is sheared randomly across the genome rather than amplified from a specific location, and thus the genetic region of a read is unknown in addition to the organism from which it came.
Because metagenomes do not proceed through an amplification step, they do not have the same PCR primer biases as a marker gene survey, however extraction efficiency concerns remain and multiplex sequencing is known to have biases of its own.

It is possible to subset metagenomic data to marker genes.
That is, one can use 16S reads that appear in the metagenome as well as reads from other ``core" genes that are expected to follow the same evolutionary path and are present in a large proportion of micro-organisms.
This is proven to be a useful strategy, and several groups have built databases of core gene families as well as provided programs and/or web tools to phylogenetically analyze metagenomes subset to those core genes \citep{von2007quantitative,wu2008amphora,stark2010mltreemap,kembel2011phylogenetic,Darling2014-ux}.
However, because of the variability of gene repertoire in microbes, this core gene set may be relatively small: even the largest collection of genes in these databases only recruits around 1 percent of a metagenome.
At least some portion of the rest of the other approximately 99 percent of the metagenome can be taxonomically classified using one of the methods described below.

Metagenomic data sets are often used to infer information about metabolism rather than phylogenetic nature \citep{abubucker2012metabolic,greenblum2012metagenomic}.
Discussing these methods is beyond the scope of this paper, as is the sequencing of mRNA in bulk which is called ``metatranscriptomics".

\subsection{Whole genomes}
In addition to conventional genome sequencing for microbial genomes, whole-genome sequencing from culture is currently being used for microbial outbreak tracking \citep{koser2012rapid,snitkin2012tracking}.
The Food and Drug Administration maintains GenomeTrakr, an openly accessible database of whole genomes sampled from the environment and grown in culture\footnote{\url{http://www.fda.gov/Food/FoodScienceResearch/WholeGenomeSequencingProgramWGS/}}.
This data may become more common for unculturable organisms as single-cell sequencing methods improve \citep[reviewed in][]{kalisky2011single}.
The assembly of complete genomes from metagenomes, once limited to samples with a very small number of organisms \citep{baker2010enigmatic}, is now becoming feasible for more diverse populations with improved sequencing technology and computational approaches \citep{emerson2012metagenomic,howe2012assembling,iverson2012untangling,pell2012scaling,podell2013assembly}.

\section{Tree-thinking in human microbiome research}

In this section I consider the ways in which phylogenetic methodology has impacted human microbiome research.
What may be most interesting for the \textit{Systematic Biology} audience is the way in which phylogenetic trees are being used to actively revise taxonomy as well as being used as a structure on which to perform sample comparison.

Before proceeding, we note that phylogenetics for microbes differs in some important respects from, say, mammalian phylogenetics.
Where phylogenetic research on mammals is converging on one or two possible basic structures for their evolution, much of the deep history of microbes remains obscure.
Inference of this history is complicated by the fact that any tree-like signal in the deep evolutionary relationships between bacteria is restricted to a small set of so-called ``core'' genes \citep{bapteste2009prokaryotic,Leigh2011-zu,Lang2013-zt}.

\subsection{Phylogenetics and taxonomy}

Phylogenetic inference has had a substantial impact on microbial ecology research by changing our view of the taxonomic relationships between microorganisms.
The clearest such example is the discovery that archaea, although similar to bacteria in their gross morphology, form their own separate lineage \citep{woese1977phylogenetic}.

Several groups are continually revising taxonomy using the results of phylogenetic tree inference.
These attempts are less ambitious than the PhyloCode project \citep[to develop a taxonomic scheme expressed directly in terms of a phylogeny; see][]{forey2001phylocode}, and simply work to revise the hierarchical structure of the taxonomy while for the most part leaving taxonomic names fixed.
Bergey's Manual of Systematic Bacteriology has officially adopted 16S as the basis for their taxonomy \citep{kreig1984bergey}, although the actual revision process appears opaque.
The GreenGenes group \citep{desantis2006greengenes} has been very active in updating their taxonomy according to 16S, first with their GRUNT tool \citep{dalevi2007automated} and more recently with their tax2tree tool \citep{mcdonald2011improved}.
Tax2tree uses a heuristic algorithm to reassign sequence taxonomic labels so that they are concordant with a given rooted phylogenetic tree in a way that allows polyphyletic taxonomic groups.
\citet{matsen2012reconciling} developed algorithms to quantify discordance between phylogeny and taxonomy based on a coloring problem previously described in the computer science literature \citep{moran2008convex}.
Although it is wonderful that several groups are actively working on taxonomic revision, it can be frustrating to have multiple different taxonomies with no easy way to translate between them or to the taxonomic names provided in the NCBI or EMBL sequence databases.

An obvious application of phylogenetics is to perform taxonomic classification, as the taxonomy is at least in part defined by phylogeny.
However, comparisons of taxonomic classification programs \citep{liu2008accurate,bazinet2012comparative} have indicated that current implementations of phylogenetic methods do not perform as well as simple classifiers based on counts of DNA substrings of a given length, which are often called $k$-mer classifiers \citep{wang2007naive,rosen2008metagenome}.
For those studies and others, the conventional metrics of classification performance such as precision and recall are applied to data simulated from known and taxonomically classified genomes.
Some authors report that a combination of composition-based and homology-based classifiers work best \citep{brady2009phymm,parks2011classifying}.
The MEGAN program \citep{huson2007megan,huson2011integrative} BLASTs an unknown sequence onto a database of sequences with taxonomic labels and assigns the sequence the lowest (i.e.\ narrowest) taxonomic group shared by all of the high-quality hits; as such it is somewhat phylogenetic in that it uses the structure of the taxonomic tree.
\citet{munch2008fast,munch2008statistical} infer taxonomic assignment by automatically retrieving sequences equipped with taxonomic information and building a tree on them along with an unknown sequence.
\citet{srinivasan2012bacterial} find that phylogenetic methods to do taxonomic classification can outperform composition-based techniques at least for certain taxonomic groups.
\citet{segata2012metagenomic} propose a clever approach to inferring organisms present in a metagenomic sample by compiling a database of clade-specific genes, then classifying a given read as being from the only clade that has the corresponding gene.
They show that this has good sensitivity and specificity, however, this method can only be used to identify the presence of organisms whose genome has been sequenced.
Other metagenomic classification techniques have been reviewed by \citet{mande2012classification}; interesting recent progress has been made by \citet{Lanzen2012-ul}, \citet{Koslicki2013-mc}, and \citet{Droge2014-lj}.
\citet{treangen2013metamos} report shorter run times and much higher accuracy (again, for taxonomic classification using reads simulated from full genomes) for such metagenomic classification when reads are assembled before they are classified.

\subsection{The role of Operational Taxonomic Units (OTUs)}
Although there continues to be a lively debate on if there is a meaningful concept of species for microbes \citep{bapteste2009prokaryotic,caro2012bacterial}, a substantial part of human microbiome research has replaced any traditional species concept with the notion of Operational Taxonomic Units (OTUs).
An OTU is a proxy species concept that is typically defined with a fixed divergence cutoff, most commonly at 97\% sequence identity, such that each OTU is a cluster of sequences that are closer to each other than that cutoff.
It is common for trees to be built on sequence representatives from these OTUs, and the abundance of an OTU to be given by the number of sequences that sit within that cluster.
I briefly describe the mini-industry of OTU clustering techniques to contrast with the phylogenetic literature on species delimitation \citep{pons2006sequence,yang2010bayesian}.
I will use the term \emph{OTU inference} despite the fact that there is no clear definition of the OTU concept.

It is not straightforward to define a clear notion of optimality for OTU inference.
While in phylogenetics we would like to compare reconstructions to an object that is generally not knowable---the ``true'' historical phylogenetic tree---OTU inference has a variety of desirable outcomes, only some of which are knowable.
\citet{Wang2013-ol} divide performance measures into \textit{external} measures, which compare an inferred clustering to some defined outcome, and \textit{internal} measures, which give overall descriptive statistics on an inferred clustering.
External measures applied to observed or simulated data include obtaining the same number of OTUs as taxonomic groups at some level \citep[e.g.][]{edgar2013uparse}, scoring deviation from a decomposition of a phylogenetic tree \citep[e.g.][]{navlakha2009finding}, or a given set of taxonomic classifications \citep[e.g.][]{Cai2011-bi}.
Internal measures evaluate the clustering in various ways without reference to a ``true'' clustering, with the general idea that within-cluster distances should be small compared to between-cluster distances.

There are many OTU inference methods with various speeds and strategies.
Some methods proceed through a list of sequences and progressively add each either to an existing cluster or start a new cluster with that sequence, such as CD-HIT \citep{li2006cdhit} and USEARCH \citep{edgar2010usearch}, while UPARSE \citep{edgar2013uparse} uses a similar strategy while also attempting to correct for sequencing error.
\citet{white2010alignment} show that different ways of doing this genre of heuristic clustering can result in very different results.
\citet{Cai2011-bi} perform highly efficient clustering using a pseudometric based partition tree, which can be thought of as a hierarchical clustering tree with a fixed set of internal node heights.
\citet{navlakha2009finding} take a semi-supervised approach in that input includes sequences along with a subset of sequences that are equipped with taxonomic classifications; the algorithm then groups all sequences (many in novel clusters) into clusters that have similar properties as the example taxonomic clusters.
\citet{Wang2013-ol} optimize a criterion of cluster modularity.
\citet{hao2011clustering} use a Gaussian mixture model formulation for clustering to avoid fixed cutoff values, and \citet{CWC12} use a two step process, first with a Dirichlet multinomial mixture on 3-mer profiles, and then a minimum description length criterion.
\citet{Zhang2013-wh} have developed a phylogenetic means to do species delimitation that scales to a relatively large number of sequencing reads and so can be used as an OTU clustering method.

The centrality of the OTU concept can be seen by the fact that the by-sample table of OTU observations (i.e.\ the matrix of counts with rows representing samples and columns representing OTUs) is considered to be the fundamental data type for 16S studies \citep{mcdonald2012biological}, or that methods have been devised to find OTUs from non-overlapping sequences \citep{sharpton2011phylotu}.
A significant amount of effort has been made to distinguish sequencing error in environmental samples from true rare variants; much of this work has played out in the OTU inference literature \citep{quince2009accurate,quince2011removing,bragg2012fast,edgar2013uparse} as such errors are especially problematic there.
With the exception of the work of \citet{sharpton2011phylotu} and \citet{Zhang2013-wh}, OTU inference is not considered to be a phylogenetic problem but rather something to be performed before phylogenetic inference begins.

\subsection{Diversity estimates using phylogenetics}

Because 16S surveys are inherently complex and noisy data, summary statistics are often used; summaries of the diversity of a single sample are often called \emph{alpha diversity}.
For the most part, this literature adapts methods from the classical ecological literature by substituting OTUs for taxonomic groups.
For example, the most commonly used index is the Simpson index \citep{simpson1949measurement}, which is simply the sum of the squared frequencies of the OTUs.
The drawback of applying such a diversity metric to a collection of OTUs is that the large-scale structure of the diversity is lost, such that two closely related OTUs contribute as much to the diversity measure as two distant ones.
Phylogenetic diversity metrics, which do take this overall diversity structure into account, are also used.
However, whereas just about every one of the hundreds or thousands of 16S surveys applies an OTU-based alpha diversity estimate, only a few involve PD.

\newcommand{\pdLegend}{
    Unweighted phylogenetic diversity (PD, left) and an abundance-weighted PD measure (right), where taxa present in a sample are shown as circles and abundances are shown as the size of the circles.
    Unweighted PD takes the total length of branches sitting between tree tips represented in a sample.
    Abundance-weighted measures take a weighted sum of branch lengths where weight is determined in some way by the abundance of the taxa on either side of the branch: if we give edges width according to their weight, the abundance-weighted measure can be thought of as the sum of the total area of the edges.
    One such abundance-weighted measure simply takes the absolute value of the difference of the total read abundance on one side compared to the other.
}
\forarxiv{
\begin{wrapfigure}[27]{r}{2.2in}
  \begin{center}
    \includegraphics[width=2.2in]{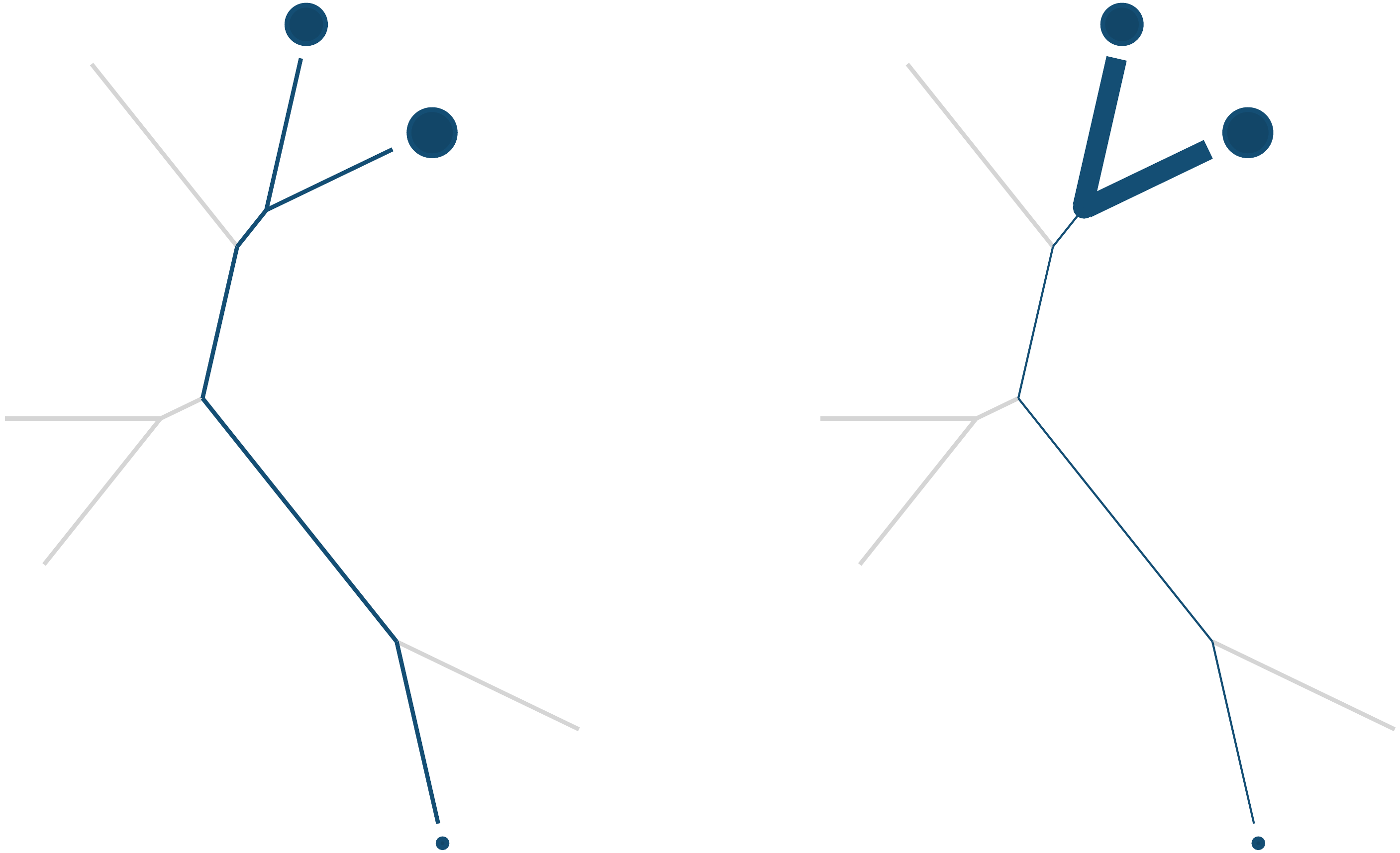}
  \end{center}
  \vspace{-10pt}
  \caption{\pdLegend}
  \label{fig:pd}
\end{wrapfigure}
}

Phylogenetic diversity (PD) measures use the structure and branch lengths of a phylogenetic tree to quantify the diversity of a sample (Fig.~1).
The (unweighted) PD of a set of taxa $S$ in an unrooted phylogenetic tree is simply the total length of the branches that sit between taxa in $S$ \citep{faith1992conservation}.
It quantifies the ``amount of evolution'' contained in the evolutionary history of those taxa.
Unweighted PD has been applied to some 16S survey data \citep{lozupone2007global,costello2009bacterial} and to metagenomic reads in a set of marker genes \citep{kembel2011phylogenetic}.

Although abundance weighted non-phylogenetic diversity measures such as \citet{simpson1949measurement} and \citet{shannon1948mathematical} are commonly used in human microbiome studies, abundance weighted phylogenetic diversity measures are not.
Abundance-weighted measures take a sum of branch lengths weighted by abundance, such that branches that connect abundant taxa get a higher weight than ones that do not (Fig.~\ref{fig:pd}).
Thus rare taxa and artifactual sequences are down-weighted compared to abundant taxa.
Such measures do exist \citep{rao1982diversity,barker2002phylogenetic,allen2009new,chao2010phylogenetic,vellend2011measuring}.
Abundance-weighted measures commonly weight edges in proportion to abundance, but one can also construct ``partially abundance weighted'' measures by weighting edges by abundances transformed by a sublinear function.
\citet{mccoy2013abundance} have recently shown that such partially abundance weighted diversity measures do a good job of distinguishing between dysbiotic and ``normal" states of the human microbiota; in particular that they do a better job than the commonly-used OTU-based measures.
\citet{nipperess2013mean} have also determined formulas for the expectation and variance of PD under random sub-sampling (see discussion of rarefaction below).

\subsection{Community comparison using phylogenetics}

The level of similarity between samples or groups of samples is called \emph{beta diversity}.
As with alpha diversity, it is not uncommon to use classical measures \citep[e.g.][]{jaccard1908nouvelles} applied to OTU counts, however phylogenetics-based methods are the most popular.
They are generally variants of the ``UniFrac'' phylogenetic dissimilarity measure \citep[as described and named by][]{LozuponeKnightUniFrac05}.
\citet{kuczynski2010microbial} claim that the UniFrac framework is superior to other methods for community comparison via real data and simulations \citep[for a contrary viewpoint using simulations see][]{schloss2008evaluating}.

\newcommand{\unifracLegend}{
    The UniFrac divergence measure \citep[figure adapted from][]{LozuponeKnightUniFrac05}.
    Assume that the sequence data to build the phylogenetic tree derives from two samples: the light-shaded sample and the dark-shaded sample (green and blue in the online version).
    When the samples are interspersed across the tree (left tree), they have a smaller \textit{fraction} of branch length that sits ancestral to clades that are \textit{uniquely} composed of one sample or another, compared to when they are separate (right tree).
    The bottom pictorial equation shows the ratios of interest for UniFrac: the branch length unique to one sample divided by the total branch length.
    The ratio is smaller when the samples are interspersed (left) than they are when separate (right tree).
}
\forarxiv{
\begin{wrapfigure}[33]{r}{2.2in}
  \vspace{-4pt}
  \begin{center}
    \includegraphics[width=2.2in]{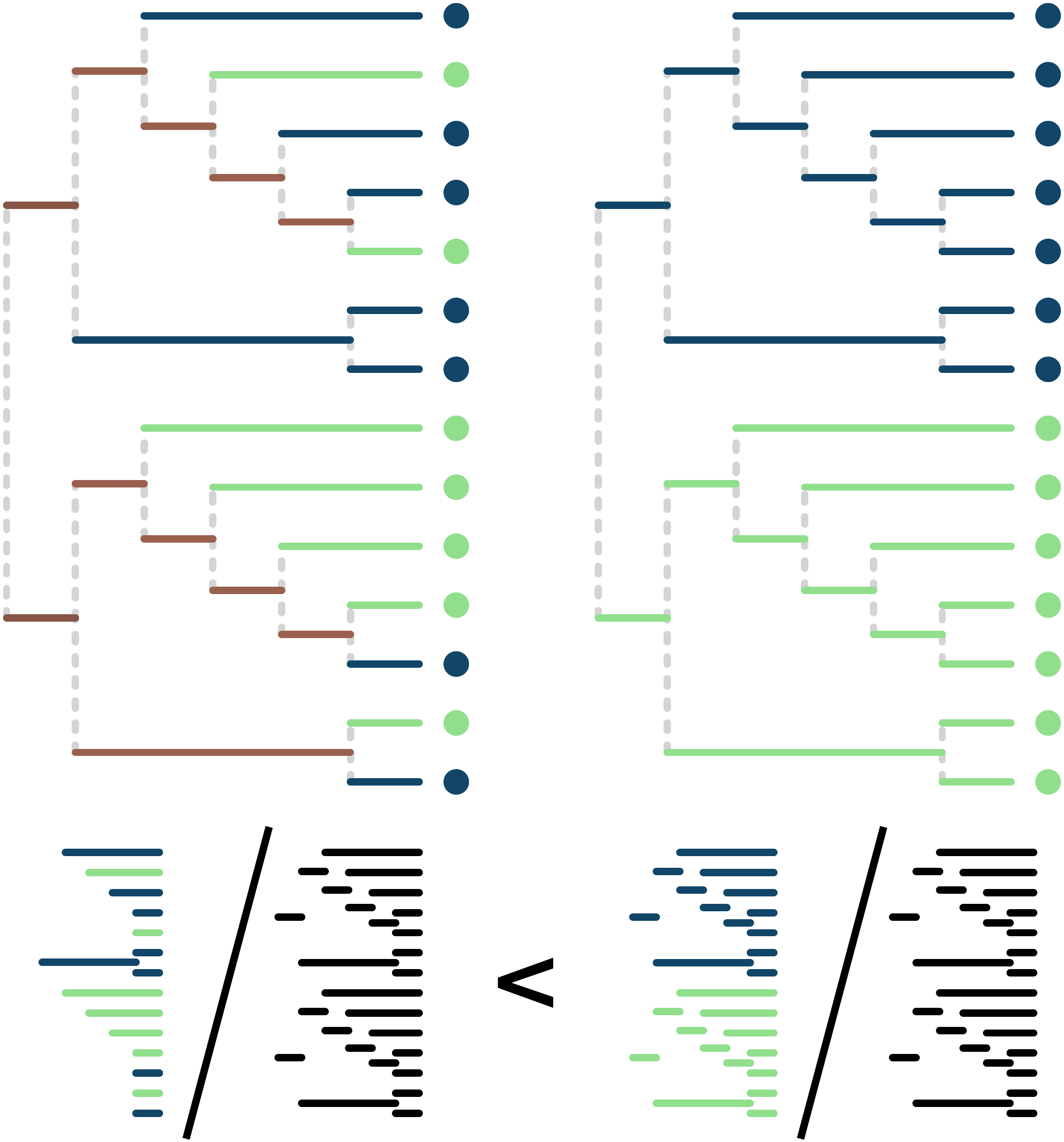}
  \end{center}
  \vspace{-10pt}
  \caption{\unifracLegend}
  \label{fig:unifrac}
\end{wrapfigure}
}

To calculate the UniFrac divergence between two samples of reads, one builds a tree on both samples and then performs a calculation on the branch lengths according whether branches have descendants in both samples or one sample only.
Specifically, the original UniFrac divergence is the fraction of the total branch length of the tree coming from branches with descendant leaves in exactly one of the two samples \citep[Fig.~\ref{fig:unifrac};][]{LozuponeKnightUniFrac05}.
Weighted UniFrac is an abundance weighted version \citep{LozuponeEaWeightedUnifrac07}.
These dissimilarity measures have hundreds of citations.
\citet{evans2012phylogenetic} showed that weighted UniFrac is in fact a specific case of the earth-movers distance, and that the commonly-used randomization procedure for significance estimation has a central limit theorem approximation.
The earth-movers distance \citep{monge1781memoire,Villani2003-wv} between two probability distributions in this case can be defined using a physical analogy as the minimum amount of ``work'', defined as mass times distance, required to move probability mass in one distribution to another along the tree.
In this case the size of the dirt piles is proportional to the number of reads mapping to that location in the tree (Fig.~\ref{fig:dirtpiles}).
\citet{chen2012associating} have shown that a partially abundance-weighted variant of UniFrac may have greater power to resolve community differences than either unweighted or weighted UniFrac.
There are clear connections between UniFrac and phylogenetic diversity explored by \citet{faith2009cladistic}, who point out related measures in Faith's earlier work.

The most common way to use a distance matrix obtained from applying UniFrac to all pairs of samples is to apply an ordination method such as principal coordinates analysis.
Indeed, the separation of two communities in a principal components plot is often used as \emph{prima facie} evidence of a difference between them
\citep[e.g.][]{lozupone2007global,costello2009bacterial,yatsunenko2012human}, while the lack of such a difference is interpreted as showing that the communities are not different overall.

There have been several efforts to augment these ordination visualizations with additional information giving more structure to the visualizations.
Biplots display variables (in the microbial case summarized by taxonomic labels) as points along with the points representing samples \citep[e.g.][]{hewitt2013bacterial,lozupone2013meta}.
\citet{PurdomAnalyzingDataGraphs08} describes how generalized principal component eigenvectors can be interpreted via weightings on the leaves of a phylogenetic tree.
\citet{matsen2013edge} have developed a variant of principal components analysis that explicitly labels the axes with weightings on phylogenetic trees that indicate their influence.

\newcommand{\dirtpilesLegend}{
    Part of a minimal mass movement to calculate the earth-mover's distance between two probability distributions on a phylogenetic tree.
    For this, each probability distribution is considered as a configuration of dirt piles (round bumps in the figure) on the tree, and the distance between two such dirt pile configurations is defined to be the minimum amount of physical ``work'' required to move the dirt in one configuration to the other.
}
\forarxiv{
\begin{wrapfigure}[25]{l}{2.in}
  \vspace{-17pt}
  \begin{center}
    \includegraphics[width=1.85in]{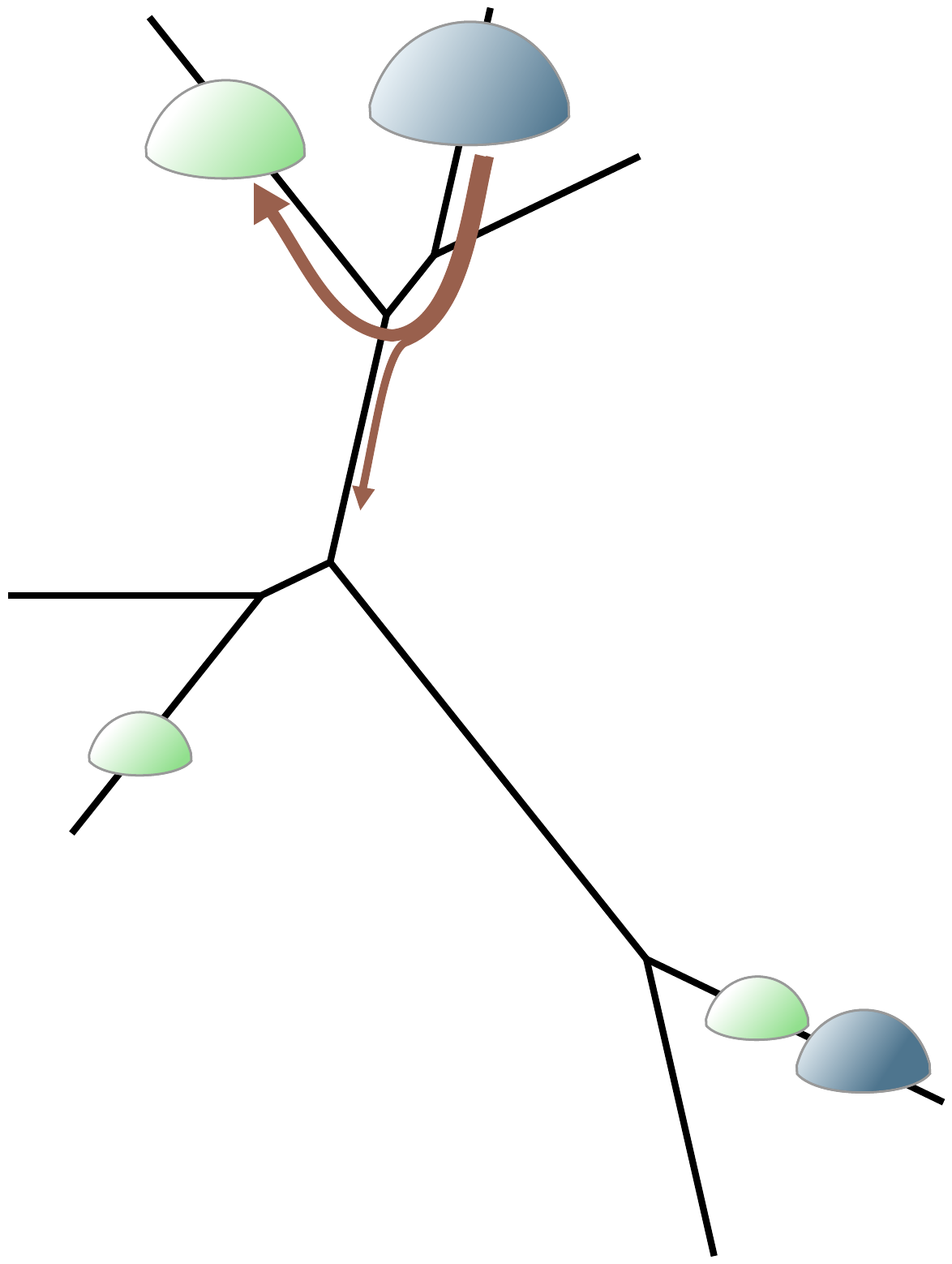}
  \end{center}
  \vspace{-5pt}
  \caption{\dirtpilesLegend}
  \label{fig:dirtpiles}
\end{wrapfigure}
}

In another vein, \citet{la2012statistical} consider the induced taxonomic tree of a sample as a statistical object and, using a framework where a sampling probability is defined in terms of a distance between such induced trees, define and investigate maximum likelihood estimation of and likelihood ratio tests for these trees.
They focus on distances between trees induced by matrix metrics on the corresponding adjacency matrices.
A similar framework was used by \citet{steel2008maximum} to construct maximum likelihood supertrees, for which they use common distance measures in phylogenetics such as the subtree-prune-regraft metric.

\subsection{Phylogeny and function}

16S distance is frequently used as a proxy for a functional comparison between human microbiome samples.
Indeed, researchers using UniFrac don't always think of their comparisons as being in terms of a single gene, but rather in terms of an abstracted measure of community function.
Those accustomed to microbial genetics may think this surprising, because the genetic repertoire of microbes is commonly acquired horizontally as well as vertically, and horizontal transmission leaves no trace in 16S ancestry.

However, \citet{zaneveld2010ribosomal} have shown that organisms that are more distant in terms of 16S are also more divergent in terms of gene repertoire.
Such observations surround a fit nonlinear curve, and the extent to which they lay on the curve appears to be phylum-dependent.
This ``proxy'' approach has recently been taken to its logical conclusion by \citet{langille2013predictive}, who develop methods to infer functional characteristics from a 16S sample using discrete trait evolution models on 16S gene trees by either parsimony \citep{kluge1969quantitative} or likelihood \citep{pagel1994detecting} methods via the ape package \citep{paradis2004ape}.

Similar logic has been applied to prioritize microbes for sequencing.
\citet{wu2009phylogeny} have derived a ``phylogeny-driven genomic encyclopaedia of Bacteria and Archaea'' by selecting organisms for sequencing that are divergent from sequenced organisms.
They have recovered more novel protein families using these phylogeny-based approaches than they would have using methods organized by selecting microbes to sequence based on their taxonomic labels.
In a similar effort for the human microbiome \citep{fodor2012most}, phylogenetic results were not shown although the authors state that phylogenetic methods did give similar results to their analysis.

\subsection{Horizontal gene transfer}
With some notable exceptions, mainstream applications of phylogenetics to a collection of human-associated microbes have typically been with the idea of finding ``the'' tree of such a collection rather than explicitly exploring divergence between various gene trees.
As described above, whole-genome data sets are typically used to directly infer functional information rather than information concerning ancestry.
The continuing debate concerning whether a microbial tree of life is a useful concept \citep{bapteste2009prokaryotic,caro2012bacterial} does not seem to have dampened human microbiome researchers' enthusiasm for using a single such tree.

Nevertheless, the work that has been done concerning horizontal gene transfer in the human microbiome has revealed interesting results.
\citet{hehemann2010transfer} found that a seaweed gene has been transferred into a bacterium in the gut microbiota of Japanese such that individuals with this resulting microbiota are better able to digest the algae in their diet.
Following on this work, \citet{smillie2011ecology} found that the human microbiome is in fact a common location for gene transfer.
\citet{stecher2012gut} found that in a mouse model, horizontal transfer between pathogenic bacteria is blocked by commensal bacteria except for periods of gut inflammation.
Horizontal transfer of genes is inferred in these studies by finding highly similar subsequences in otherwise less related organisms.

\section{Phylogenetic inference as practiced by human microbiome researchers}

\subsection{Alignment and tree inference}
In general, human microbiome researchers are interested in quickly doing phylogenetic inference on large data sets, and are less interested in clade-level accuracy or measures of uncertainty.
This is defended by saying that for applications such as UniFrac, the tree is used as a framework to structure the data, and there is a certain amount of flexibility in that framework that will give the same results.
Furthermore, given that the underlying data sets are typically 16S alone we can expect some topological inaccuracy in reconstructing the ``tree of cells'' even with the best methods.
Additionally, as specified below, these data sets can be very large.
There does not seem to be contentious discussion of specific features of the inferred trees equivalent to, say, the current discussion around the rooting of the placental mammal tree \citep{morgan2013heterogeneous,romiguier2013less}.
Given this perspective, it is not surprising that Bayesian phylogenetic methods and methods that incorporate alignment uncertainty are absent.

Alignment methods are primarily focused on developing automated methods to extend a relatively small hand-curated ``seed alignment'' with additional sequences; several tools have been created with exactly this application for 16S in mind \citep{desantis2006nast,caporaso2010pynast,pruesse2012sina}.
The community also uses profile hidden Markov models \citep{eddy1998profile} and CM models \citep{nawrocki2009infernal,nawrocki2009structural} to achieve the same result.

The large data sets associated with human microbiome analysis require highly efficient algorithms for \emph{de novo} tree inference.
Historically this has meant relaxed neighbor joining \citep{evans2006relaxed}, but more recently FastTree 2 \citep{price2010fasttree} has emerged as the \textit{de facto} standard.
Researchers do most phylogenetic inferences as part of a pipeline such as mothur \citep{schloss2009introducing} which has incorporated the clearcut code \citep{sheneman2006clearcut}, or QIIME \citep{caporaso2010qiime}, which wraps clearcut, FastTree, and RAxML \citep{Stamatakis2006-yz}.

The scale of the data has motivated strategies other than complete phylogenetic inference, such as the insertion of sequences into an existing phylogenetic tree.
Although such insertion has long been used as a means to build a phylogenetic tree sequentially \citep{kluge1969quantitative}, the first software with insertion specifically as a goal was the parsimony insertion tool in the ARB program by \citet{ludwig2004arb}.
ARB is commonly used to reconstruct a full tree by direct insertion.

There are also other methods with the less ambitious goal of mapping sequences of unknown origin into a so-called fixed reference tree, sometimes with uncertainty estimates.
These programs \citep{monierEaLargeViruses08,vonMeringEaQuantitative08,wu2008simple,matsen2010pplacer,stark2010mltreemap,berger2011performance} have various speeds and features.
This work has also spurred development of specialized alignment tools for this mapping process.
\citet{berger2011aligning} focus on the problem of inferring the optimal alignment and insertion of sequences into a tree.
\citet{Mirarab2012-pk} use data set partitioning to improve alignments on subsets of taxa specifically for this application.
\citet{Brown2013-ia} use locality-sensitive hashing to obtain placement more than two orders of magnitude faster than the \textit{pplacer} program of \citet{matsen2010pplacer}.

Considerable effort goes to the creation of large curated alignments and phylogenetic trees on 16S.
There are two major projects to do so: one is the SILVA database \citep{pruesse2007silva,quast2013silva}, and the other is the GreenGenes database \citep{desantis2006greengenes,mcdonald2011improved}.
Because of the high rate of insertion and deletion of nucleotides in 16S, these alignments have a high percentage of gap.
Taking the length of 16S to be 1543 nucleotides,
the 479,726 sequence SILVA reference alignment version 115 is over 96\% gap,
while the 1,262,986 sequence GreenGenes 13\_5 alignment is almost 80\% gap.
The SILVA-associated `all-species living tree' project \citep{yarza2008all} started with a tree inferred by maximum likelihood and has been continually updated  by inserting sequences via parsimony.
The GreenGenes tree is updated by running FastTree from scratch for every release.
There appears to be a commonly held belief that FastTree in particular works well even with such gappy alignments \citep[e.g.][]{sharpton2011phylotu}.

In addition to these 16S-based resources, the MicrobesOnline resource \citep{dehal2010microbesonline} offers a very nice interactive tree-based genome browser.
On a much smaller scale, there are microbiome body-site specific reference sequence sets \citep{chen2010human,griffen2011core,srinivasan2012bacterial}

\section{Phylogenetic challenges and opportunities in human microbiome research}

Many phylogenetic challenges remain in human microbiome research.
Some of them are familiar, such as how to build large phylogenies on data that has many insertions and deletions.
I review some others here.

One clear challenge is to fill the gap between on one hand complete \emph{de novo} tree inference versus sequence insertion or placement that leaves the ``reference tree'' fixed, with the idea that such an algorithm would retain the efficiency characteristics of placement algorithms while allowing the reference tree to change.
For example, sequence data sets are continually being added to large databases, motivating methods that could continually update trees with this new sequence data while allowing the previous tree to change according to this new information.
\citet{Izquierdo-Carrasco2014-hu} have taken a step in this direction by developing an informatic framework that updates alignments and builds larger trees using previous smaller trees as starting points.

In this review I have devoted considerable space to the ways in which microbial ecologists have used the 16S tree as a proxy structure for the complete evolutionary history of their favorite organisms.
They have even shown that 16S distance recapitulates gene content divergence and used this correlation to predict gene functions.
It is well known, however, that any single tree will not give a complete representation of the evolutionary history of a collection of microbes.

The apparent success of 16S-tree-based comparisons raises the question of if a more complete representation of the evolutionary history of the microbes would yield better comparisons.
This suggests a practical perspective on the theoretical issue of the tree of life: what is the representation of the genetic ancestry of a set of microbes that allows us to best perform proxy whole-genome comparison?
This representation could be simple.
For example, one of the results of \citet{zaneveld2010ribosomal} is that 16S correlates better with gene repertoire in some taxonomic groups than others.
If we were to equip the 16S tree with some measure of the strength of that correlation, would that allow for more precise comparison?
If we allow an arbitrary ``hidden'' object, what such object would perform best?
\citet{Parks2012-os} have expanded the range of choices by defining community comparison metrics on phylogenetic split systems.
An alternative would be to use collections of reconciled gene trees in the presence of gene deletion, transfer, and loss \citep[e.g.][]{szollHosi2013efficient,szollHosi2013lateral}.

It appears that neutral models involving phylogenetics could be more fully developed.
Methods explicitly invoking trait evolution are notably absent, with the recent exception of the work by \citet{langille2013predictive}.
The results of this simple method are reasonable, but would a collection of gene trees reconciled with a species tree allow for better prediction?
Perhaps improved methods, say involving whole-genome evolutionary modeling or models of metabolic network evolution, could shed light on the problem.
Here again the ``tree of life'' problem can be formulated in a practical light: what representation allows for the best prediction of features of underlying genomes?
How might one formulate a useful notion of independent contrasts \citep{felsenstein1985phylogenies} on such an object?
It is quite possible that inference using a more complete representation would not be able to overcome the inherent noise of the data, but further exploration seems warranted as simple methods give reasonable results.

While developing community assembly models forms an important project for microbial ecology generally and human-associated microbial ecology in particular, phylogeny-aware methods could be further developed.
One way to model microbial community assembly is to apply Hubbell's neutral theory \citep{costello2012application,fierer2012animalcules}.
\citet{o2012phylogenetic} model community assembly with an explicitly phylogenetic perspective, and include some comparison of models to data.
Continued work in this direction seems warranted, given the way in which phylogenetic tree shape statistics have had a significant impact on macroevolutionary modeling \citep{mooers1997inferring,aldous2011five}.
In this case various (alpha and beta) diversity statistics would play the role of tree shape statistics by reducing a distribution on the tips of a tree down to a real number.
Another challenge is to bring together macroevolutionary modeling with species abundance modeling, where some initial work has been done by \citet{lambert2013predicting} in another setting.

Diversity preservation is of interest for microbiota researchers like it is for eukaryotic organisms, but has not received the formalization and algorithmic treatment surrounding phylogenetic diversity for larger organisms \citep{hartmann2006maximizing,pardi2007resource}.
Martin Blaser in particular has argued that changes in our microbiota are leading to an increase in autoimmune disease and certain types of cancer \citep[reviewed in][]{cho2012human} and has made passionate appeals to preserve microbiota diversity \citep{blaser2011antibiotic}.
Because a child's initial microbiota is transmitted from the mother \citep[reviewed in][]{funkhouser2013mom}, there is a somewhat equivalent notion of microbiota extinction when the chain is interrupted via cesarean section and infant formula.
In order to characterize extant diversity, \citet{yatsunenko2012human} have explicitly contrasted microbiota development in urban, forest-dwelling, and rural populations, while \citet{tito2012insights} have endeavored to characterize the microbiota from ancient feces.
How might phylogenetic methods be used in these preservation efforts?

There are indications of coevolution between microbiota and their hosts.
\citet{ochman2010evolutionary} found identical tree topologies for primate and microbiome evolution.
For the microbiota, they used maximum parsimony such that each column represented a microbe and each such entry took discrete states according to how much of that microbe was present.
Although parsimony gave an interesting answer here, the presence of such coevolution raises the question of what sort of forward-time models are appropriate for microbiota change?
Would methods using these models do better than parsimony or commonly-applied phenetic methods applied to the distances described above?
Some studies \citep[e.g.][]{phillips2012microbiome,delsuc2013convergence} see a combination of historical and dietary influences.
How can such forces be compared in this setting?

As described above, \citet{morgan2010metagenomic} showed that various microbes have different DNA extraction efficiencies, meaning that the representation of marker gene sequences is not representative of the actual communities.
Furthermore, there was no clear taxonomic signal in their observations of the variability of extraction efficiency, which seems to preclude a correction strategy based on ``species-tree'' phylogenetic modeling.
However, presumably \textit{something} about their genome is determining extraction efficiency; it would be interesting and useful to search for the genetic determinants.
As described above, abundances are commonly used as part of community comparison, thus a better quantification of error in those observations of abundance would be a great help.

In a similar vein, assessing the significance level of an observed difference between communities poses difficult problems.
The randomization of group membership commonly used in combination with UniFrac to determine significance does not have appropriate properties in the regime of incomplete sampling with non-independent observations, which is certainly the correct regime for surveys and metagenomes.
Such non-independence can lead to incorrect rejection of the null hypothesis.
Imagine, for example, that we have a random process as follows.
Each sample from the process takes a random subset of ``observations'' from the leaves and then throws down some number of reads for each observation in that subset, with the number of reads having a mean significantly greater than 1.
If the number of leaves is large compared to the number of sample observations, then two draws will always appear significantly different even though they are from the same underlying process.
In trying to remedy such false-positive identification of differences, it becomes clear that even basic definitions pose a challenge: the question of whether two communities are the ``same'' and ``different'' probably needs to be approached from the perspective of ecosystem modeling.

For both alpha and beta diversity measurement, read count normalization has not received nearly the attention that it has in other applications of high throughput sequencing \citep[such as RNA-Seq, e.g.][]{anders2010differential,robinson2010edger}.
One type of normalization handles differential depths of sequencing across samples.
The presently used approach is \textit{rarefaction}, which means uniform sub-sampling to the number of reads in the lowest abundance sample \citep{schloss2009introducing,caporaso2010qiime}.
In addition to throwing away data, this normalization implicitly assumes a model whereby reads are sequenced independently of one another.
This is not the case.
An alternative is provided by \citet{o2012phylogenetic}, who provide a ``UniFrac score normalization curve'' based on a sampling model of community assembly.
This is a good start, but more work should be done exploring results under deviation from that model.

Another type of normalization seeks to infer the true abundances from noisy observations of the various taxonomic groups or OTUs.
\citet{holmes2012dirichlet} and \citet{la2012hypothesis} use models where read counts are modeled as overdispersed samples of the true abundance and provide methods for statistical testing.
\citet{paulson2013differential} estimate true abundances using a zero-inflated Gaussian mixture model for read counts, while \citet{McMurdie2014-mn} claim better performance using a Gamma-Poisson mixture.

This work could be extended to a phylogenetic context by making use of the relationship between OTUs, and modeling the way in which the abundance of one OTU may increase the abundance of a related OTU because of sequencing error or a change of condition that changes the abundance of both.

Finally, the conventional wisdom that UniFrac analysis is robust to tree reconstruction methodology begs further exploration.
Would it be possible to infer an equivalence class of phylogenetic trees, where two trees are deemed equivalent if they induce the same principal coordinates projection given the same underlying presence/absence or count data?
Given that a tree is an integral part of a UniFrac analysis, it would be interesting to be able to infer the features of a tree that determine the primary trends in a projection.

\section{Discussion}
What can we expect next at the intersection of phylogenetics and the human microbiota?
At least for the next several years we can expect the research questions described above to continue to unfold.
Future research projects will continue to bring deeper sequencing on more samples.
The uBiome\footnote{\url{http://ubiome.com/}} and American Gut\footnote{\url{http://americangut.org/}} projects promise to bring gut microbiome sequencing to the average citizen for a low price.
Comparative studies will continue to investigate what shapes and is shaped by the microbiota.
However, some of the initial excitement may have died down, as neither the Human Microbiome Project nor the MetaHIT project were extended.

There are limitations to what we can learn using genetic sequences because more intricate processes such as gene regulation may be at play, limiting what sequence-level phylogenetics can do.
Future work may move from general ecological models to models that include specific interactions between microbes and the host \citep[reviewed in][]{hooper2012interactions}.

Opportunities for clinical applications will present themselves, although the specifics will change.
For example, routine 16S sequencing is likely to be replaced soon by Matrix-Assisted Laser Desorption Ionization--Time of Flight (MALDI-TOF) mass spectrometry for assignment of a single microbe grown in culture to a database entry \citep{clark2013matrix}.
However, for diagnoses that are on the level of microbial communities, sequencing and consequent analysis methods (possibly including phylogenetics) will still be required \citep[reviewed in][]{Rogers2013271}.
Inexpensive whole-genome sequencing will certainly have a profound impact on clinical practice and epidemiological studies \citep{didelot2012transforming}, and this genome-scale data requires evolutionary analysis methods to interpret it.
For all of these measures, it will be important to have rigorous means of quantifying uncertainty for robust diagnostic applications.

Human microbiome research has experienced a frenetic rate of expansion over the past decade, and sometimes the hype has outmatched the science.
However, our microbes are here to stay and so is research on them.
Thus we can look forward to the field of human microbiome analysis settling down to a comfortable and mature middle age as an interesting intersection between ecology and medicine.
Phylogenetics has already contributed significantly to research on the human microbiome and will continue to do so.

\section{Acknowledgements}
I thank Olivier Gascuel for the opportunity to present on this subject during the 2013 MCEB Mathematical and Computational Evolutionary Biology workshop and for organizing a corresponding special section of \textit{Systematic Biology}.
I am grateful to Aaron Darling, David Fredricks, Noah Hoffman, Steven Kembel, Connor McCoy, Martin Morgan, and Sujatha Srinivasan for interesting discussions that informed this review, thank Bastien Boussau, Noah Hoffman, Christopher Small and Bj\"orn Winckler for providing feedback on the manuscript.
The manuscript was greatly improved by thoughtful peer review from Frank (Andy) Anderson, Olivier Gascuel, Alexis Stamatakis, Frédéric Delsuc, and an anonymous referee.
This work was supported by National Institutes of health grant R01-HG005966-01 and National Science Foundation grant 1223057.

\notforarxiv{
\newpage
\section{Figure Legends}

\noindent
\begin{figure}[ht]
  \caption{\pdLegend}
  \label{fig:pd}
\end{figure}

\noindent
\begin{figure}[ht]
  \caption{\unifracLegend}
  \label{fig:unifrac}
\end{figure}

\noindent
\begin{figure}[ht]
  \caption{\dirtpilesLegend}
  \label{fig:dirtpiles}
\end{figure}

\newpage
}

\bibliographystyle{sysbio}
\bibliography{sbreview}

\end{document}